\documentclass[aps,prl,showpacs,twocolumn,groupedaddress,superscriptaddress]
{revtex4-1} 
\usepackage{graphicx} 
\usepackage{dcolumn} 
\usepackage{longtable} 
\usepackage{amsmath} 

\begin{document} 

\title{Evidence for $_{\Lambda}^{6}{\rm H}$} 

\author{M.~Agnello}\affiliation{Dipartimento di Fisica, Politecnico di Torino, 
Corso Duca degli Abruzzi 24, Torino, Italy}\affiliation{INFN Sezione di 
Torino, via P. Giuria 1, Torino, Italy} \author{L.~Benussi}\affiliation
{Laboratori Nazionali di Frascati dell'INFN, via E. Fermi 40, Frascati, Italy} 
\author{M.~Bertani}\affiliation{Laboratori Nazionali di Frascati dell'INFN, 
via E. Fermi 40, Frascati, Italy} \author{H.C.~Bhang}\affiliation{Department 
of Physics, Seoul National University, 151-742 Seoul, South Korea} 
\author{G.~Bonomi}\affiliation{Dip. di Ingegneria Meccanica e Industriale, 
Universit\`a di Brescia, via Valotti 9, Brescia, Italy} 
\affiliation{INFN Sezione di Pavia, via Bassi 6, Pavia, Italy} 
\author{E.~Botta}\thanks{Corresponding author: Elena Botta, botta@to.infn.it}
\affiliation{Dipartimento di Fisica Sperimentale, Universit\`a di Torino, 
Via P. Giuria 1, Torino, Italy}
\affiliation{INFN Sezione di Torino, via P. Giuria 1, Torino, Italy} 
\author{M.~Bregant}\affiliation{SUBATECH, \'Ecole des Mines de Nantes, 
Universit\'e de Nantes, CNRS-IN2P3, Nantes, France} \author{T.~Bressani}
\affiliation{Dipartimento di Fisica Sperimentale, Universit\`a di Torino, 
Via P. Giuria 1, Torino, Italy}\affiliation{INFN Sezione di Torino, via 
P. Giuria 1, Torino, Italy} \author{S.~Bufalino}\affiliation{INFN Sezione 
di Torino, via P. Giuria 1, Torino, Italy} \author{L.~Busso}\affiliation
{Dipartimento di Fisica Generale, Universit\`a di Torino, Via P. Giuria 1, 
Torino, Italy}\affiliation{INFN Sezione di Torino, via P. Giuria 1, Torino, 
Italy} \author{D.~Calvo}\affiliation{INFN Sezione di Torino, via P. Giuria 1, 
Torino, Italy} \author{P.~Camerini}\affiliation{INFN Sezione di Trieste, via 
Valerio 2, Trieste, Italy}\affiliation{Dipartimento di Fisica, Universit\`a 
di Trieste, via Valerio 2, Trieste, Italy} \author{B.~Dalena}\affiliation
{CEA, Irfu/SACM, Gif-sur-Yvette, France} \author{F.~De Mori}\affiliation
{Dipartimento di Fisica Sperimentale, Universit\`a di Torino, Via P. Giuria 1, 
Torino, Italy}\affiliation{INFN Sezione di Torino, via P. Giuria 1, Torino, 
Italy} \author{G.~D'Erasmo}\affiliation{Dipartimento di Fisica Universit\`a 
di Bari, via Amendola 173, Bari, Italy}\affiliation{INFN Sezione di Bari, 
via Amendola 173, Bari, Italy} \author{F.L.~Fabbri}\affiliation{Laboratori 
Nazionali di Frascati dell'INFN, via E. Fermi, 40, Frascati, Italy} 
\author{A.~Feliciello}\affiliation{INFN Sezione di Torino, via P. Giuria 1, 
Torino, Italy} \author{A.~Filippi}\affiliation{INFN Sezione di Torino, via 
P. Giuria 1, Torino, Italy} \author{E.M.~Fiore}\affiliation{Dipartimento di 
Fisica Universit\`a di Bari, via Amendola 173, Bari, Italy}\affiliation{INFN 
Sezione di Bari, via Amendola 173, Bari, Italy} \author{A.~Fontana}
\affiliation{INFN Sezione di Pavia, via Bassi 6, Pavia, Italy} \author
{H.~Fujioka}\affiliation{Department of Physics, Kyoto University, Sakyo-ku, 
Kyoto, Japan} \author{P.~Genova}\affiliation{INFN Sezione di Pavia, via Bassi 
6, Pavia, Italy} \author{P.~Gianotti}\affiliation{Laboratori Nazionali di 
Frascati dell'INFN, via E. Fermi 40, Frascati, Italy} \author{N.~Grion}
\affiliation{INFN Sezione di Trieste, via Valerio 2, Trieste, Italy} 
\author{V.~Lucherini}\affiliation{Laboratori Nazionali di Frascati dell'INFN, 
via E. Fermi 40, Frascati, Italy} \author{S.~Marcello}\affiliation
{Dipartimento di Fisica Sperimentale, Universit\`a di Torino, Via P. Giuria 1, 
Torino, Italy}\affiliation{INFN Sezione di Torino, via P. Giuria 1, Torino, 
Italy} \author{N.~Mirfakhrai}\affiliation{Department of Physics, Shahid 
Behesty University, 19834 Teheran, Iran} \author{F.~Moia}\affiliation
{Dipartimento di Meccanica, Universit\`a di Brescia, via Valotti 9, Brescia, 
Italy}\affiliation{INFN Sezione di Pavia, via Bassi 6, Pavia, Italy} 
\author{O.~Morra}\affiliation{INAF-IFSI, Sezione di Torino, Corso Fiume 4, 
Torino, Italy}\affiliation{INFN Sezione di Torino, via P. Giuria 1, Torino, 
Italy} \author{T.~Nagae}\affiliation{Department of Physics, Kyoto University, 
Sakyo-ku, Kyoto, Japan} \author{H.~Outa}\affiliation{RIKEN, Wako, Saitama 
351-0198, Japan} \author{A.~Pantaleo}\thanks{deceased}\affiliation{INFN 
Sezione di Bari, via Amendola 173, Bari, Italy} \author{V.~Paticchio}
\affiliation{INFN Sezione di Bari, via Amendola 173, Bari, Italy} 
\author{S.~Piano}\affiliation{INFN Sezione di Trieste, via Valerio 2, Trieste, 
Italy} \author{R.~Rui}\affiliation{INFN Sezione di Trieste, via Valerio 2, 
Trieste, Italy}\affiliation{Dipartimento di Fisica, Universit\`a di Trieste, 
via Valerio 2, Trieste, Italy} \author{G.~Simonetti}\affiliation{Dipartimento 
di Fisica Universit\`a di Bari, via Amendola 173, Bari, Italy}\affiliation
{INFN Sezione di Bari, via Amendola 173, Bari, Italy} \author{R.~Wheadon}
\affiliation{INFN Sezione di Torino, via P. Giuria 1, Torino, Italy} 
\author{A.~Zenoni}\affiliation{Dipartimento di Meccanica, Universit\`a di 
Brescia, via Valotti 9, Brescia, Italy}\affiliation{INFN Sezione di Pavia, 
via Bassi 6, Pavia, Italy} \collaboration{The FINUDA Collaboration} 
\noaffiliation 
\author {A.~Gal} 
\affiliation{Racah Institute of Physics, The Hebrew University, Jerusalem 
91904, Israel} 

\date{\today} 
       
\begin{abstract} 
Evidence for the neutron-rich hypernucleus $_{\Lambda}^{6}{\rm H}$ is 
presented from the FINUDA experiment at DA$\Phi$NE, Frascati, studying 
$(\pi^+,\pi^-)$ pairs in coincidence from the $K^{-}_{\rm stop}+{^{6}{\rm Li}}
\rightarrow {^{6}_{\Lambda}{\rm H}}+\pi^{+}$ production reaction followed by 
$_{\Lambda}^{6}{\rm H}\to {^{6}{\rm He}}+\pi^-$ weak decay. The production 
rate of $_{\Lambda}^{6}{\rm H}$ undergoing this two-body $\pi^-$ decay 
is determined to be $(2.9\pm 2.0)\cdot 10^{-6}/K^{-}_{\rm stop}$. 
Its binding energy, evaluated jointly from production and decay, 
is $B_{\Lambda}({_{\Lambda}^{6}{\rm H}})=(4.0\pm 1.1)$ MeV with respect to 
$^{5}{\rm H}+\Lambda$. A systematic difference of ($0.98\pm 0.74$) MeV between 
$B_{\Lambda}$ values derived separately from decay and from production is 
tentatively assigned to the $_{\Lambda}^{6}{\rm H}$ $0^{+}_{\rm g.s.}\to 1^+$ 
excitation. 
\end{abstract} 

\pacs{21.80.+a, 25.80.Nv, 21.10.Gv} 

\maketitle

{\textbf{Introduction.}}~
The existence and observability of neutron-rich $\Lambda$ hypernuclei was 
discussed back in 1963 by Dalitz and Levi-Setti \cite{DLS63} who predicted 
the stability of $_{\Lambda}^{6}{\rm H}$ consisting of four neutrons, one 
proton and one $\Lambda$ hyperon. Accordingly, the $\Lambda$ hyperon 
stabilizes the core nucleus $^{5}{\rm H}$ which is a broad resonance 1.7 MeV 
above $^{3}{\rm H}+2n$ \cite{Kor01}. To be stable, $_{\Lambda}^{6}{\rm H}$ 
must lie also below $_{\Lambda}^{4}{\rm H}+2n$ which provides the lowest 
particle stability threshold. This motivates a $_{\Lambda}^{4}{\rm H}+2n$ 
two-neutron halo cluster structure for $_{\Lambda}^{6}{\rm H}$, with binding 
energy and excitation spectrum that might deviate substantially from the 
extrapolation practised in Ref.~\cite{DLS63}. Specifically, the study of 
$_{\Lambda}^{6}{\rm H}$ and of heavier neutron-rich hypernuclei that go 
appreciably beyond the neutron stability drip line in nuclear systems could 
place valuable constraints on the size of coherent $\Lambda N-\Sigma N$ 
mixing in dense strange neutron-rich matter \cite{Aka99}. This mixing provides 
a robust mechanism for generating three-body $\Lambda NN$ interactions, with 
immediate impact on the stiffness/softness of the equation of state for 
hyperons in neutron-star matter, as reviewed recently in Ref.~\cite{Sch08}. 

In this Letter we report on a study of $_{\Lambda}^{6}{\rm H}$ 
in the double charge exchange reaction at rest 
\begin{equation} 
K^{-}_{\rm stop} + {^{6}{\rm Li}} \rightarrow { _{\Lambda}^{6}{\rm H}} + \pi^+ 
\;\;\; (p_{\pi^+} \sim 252~{\rm MeV/c}) 
\label{eq:production} 
\end{equation} 
based on analyzing the total data sample of the FINUDA experiment during 
2003--2007 and corresponding to a total integrated luminosity of 1156 
pb$^{-1}$. A first analysis of partial data, corresponding to an integrated 
luminosity 190 pb$^{-1}$, gave only an upper limit for (\ref{eq:production}): 
$(2.5\pm 0.4_{\rm stat}\hspace{1mm}^{+0.4}_{-0.1{\rm syst}})\cdot 10^{-5}/
K^{-}_{\rm stop}$ \cite{FIN06}. Although the statistics collected on $^6$Li 
targets is improved by a factor five with respect to the run of the earlier 
search, the inclusive $\pi^+$ spectra do not show any clear peak attributable 
to $_{\Lambda}^{6}{\rm H}$ near $p_{\pi^+} \sim 252$~MeV/c. Exploiting the 
increased statistics, the essential idea of the present analysis was to reduce 
the overwhelming background events in reaction (\ref{eq:production}) by 
requiring coincidence with $\pi^-$ mesons from the two-body weak decay 
\begin{equation} 
_{\Lambda}^{6}{\rm H} \rightarrow {^{6}{\rm He}} + \pi^- 
\;\;\; (p_{\pi^-} \sim 134~{\rm MeV/c}),  
\label{eq:decay} 
\end{equation} 
with a branching ratio of about $50\%$ considering the value measured 
for $_{\Lambda}^{4}{\rm H} \rightarrow {^{4}{\rm He}} + \pi^-$ \cite{KEK89}. 
The analysis described below yielded three distinct $_{\Lambda}^{6}{\rm H}$ 
candidate events which give evidence for a particle-stable $_{\Lambda}^{6}$H 
with some indication of its excitation spectrum. 
The deduced $_{\Lambda}^{6}{\rm H}$ binding energy does not 
confirm the large effects conjectured in Ref.~\cite{Aka99}.

{\textbf{Data analysis.}}~ 
We first recall the experimental features relevant to the present analysis. 
For $\pi^{+}$ with momentum $\sim 250$ MeV/c the resolution of the 
tracker was determined by means of the peak due to monochromatic (236.5 MeV/c) 
$\mu^{+}$ from $K_{\mu 2}$ decay and is $\sigma _{p} = (1.1\pm 0.1)$ MeV/c 
\cite{spectrFND}, the precision on the absolute momentum calibration is 
better than 0.12 MeV/c for the $^{6}$Li targets, which corresponds to 
a systematic deviation on the kinetic energy $\sigma_{T{\rm syst}}(\pi^{+}) = 
0.1$ MeV. For $\pi^{-}$ with momentum $\sim 130$ MeV/c the resolution and 
absolute calibration were evaluated from the peak due to monochromatic (132.8 
MeV/c) $\pi^{-}$ coming from the two-body weak decay of $^{4}_{\Lambda}$H, 
produced as hyperfragment with a formation probability about $10^{-3}-10^{-2}$ 
per stopped $K^{-}$ \cite{KEK89}. A resolution $\sigma_{p}=(1.2\pm 0.1)$ MeV/c 
and precision of 0.2 MeV/c were found, corresponding to a systematic deviation 
of the kinetic energy $\sigma_{T {\rm syst}}(\pi^{-})=0.14$ MeV. 

Since the stopping time of $^{6}_{\Lambda}$H in metallic Li is shorter than 
its lifetime, both production (\ref{eq:production}) and decay (\ref{eq:decay}) 
occur at rest, and a straightforward algebra leads to the following expression 
for $T_{\rm sum}\equiv T(\pi^{+})+T(\pi^{-})$: 
\begin{eqnarray} 
T_{\rm sum} & = & M(K^{-}) + M(p) - M(n) - 2M(\pi)  \nonumber \\   
 & - & B({^{6}{\rm Li}})+B({^{6}{\rm He}})-T({^{6}{\rm He}})-T({^{6}_{\Lambda}
{\rm H}}), 
\label{Ebal} 
\end{eqnarray} 
in which $M$ stands for known masses, $B$ for known nuclear binding energies, 
and $T$ for kinetic energies. The evaluation of $T({^{6}_{\Lambda}{\rm H}})$ 
using momentum and energy conservation depends explicitly on the knowledge of 
$B_{\Lambda}({_{\Lambda}^{6}{\rm H}})$, whereas $T({^{6}{\rm He}})$ depends 
only implicitly on $B_{\Lambda}({_{\Lambda}^{6}{\rm H}})$ through the momentum 
$p_{\pi^-}$. 

\begin{figure}[htbp] 
\begin{center} 
\includegraphics[width=80mm]{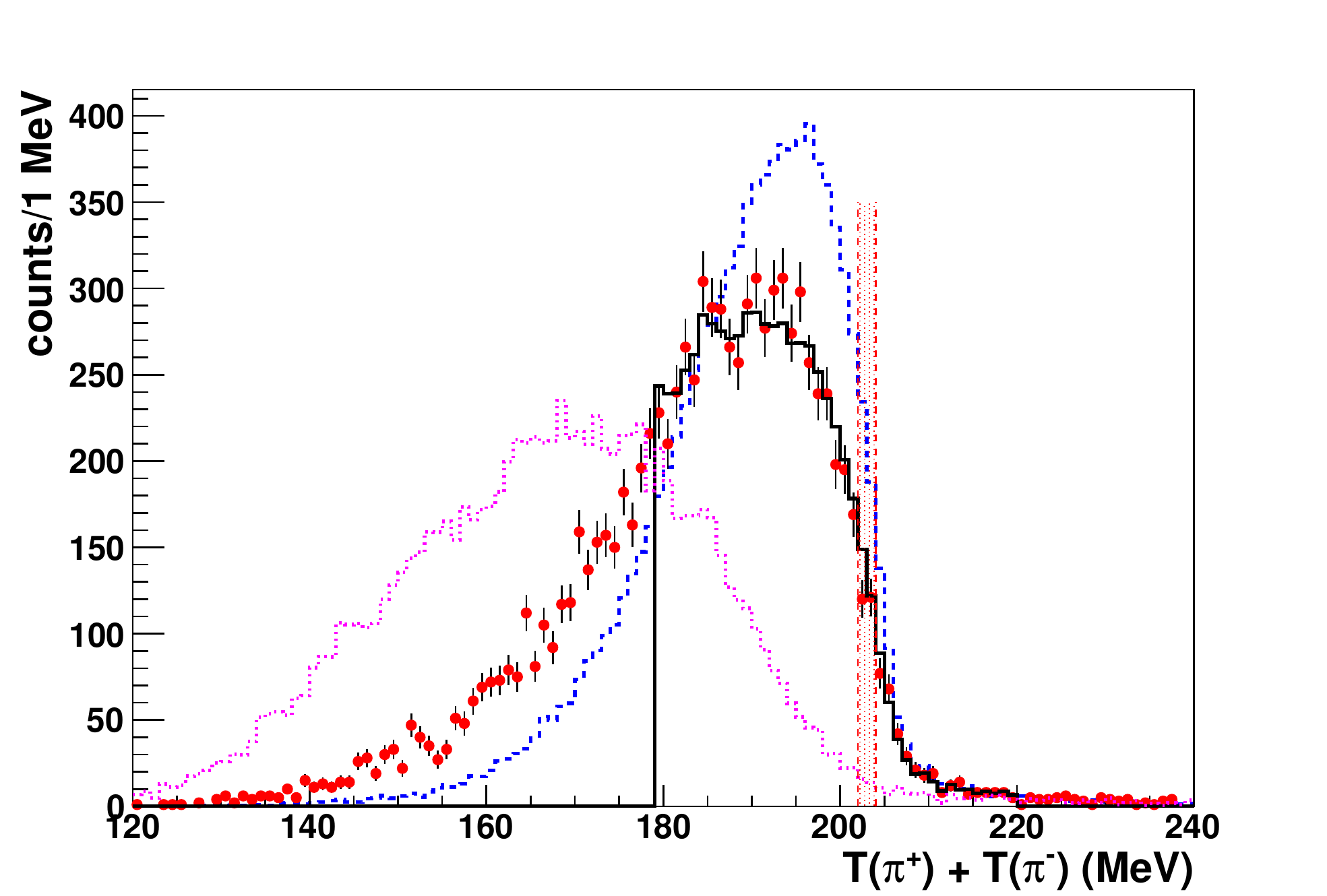} 
\caption{(color online). Distribution of raw total kinetic energy $T_{\rm sum}
\equiv T(\pi^{+})+T(\pi^{-})$ for $\pi^{\pm}$ pair coincidence events from 
$^{6}$Li targets. The vertical (red) bar represents the cut $T_{\rm sum}
=202-204$ MeV. The dashed (blue) histogram is a quasi-free simulation of  
$K^{-}_{\rm stop}+{^{6}{\rm Li}}\rightarrow\Sigma^{+}+{^{4}{\rm He}}+n+\pi^{-};
\Sigma^{+}\rightarrow n+\pi^{+}$ background and the dotted (violet) histogram 
is a four-body phase space simulation of the same background, their best fit 
to the data is shown by the solid (black) histogram, see text. 
}
\label{fig3} 
\end{center} 
\end{figure}

We assume $B_{\Lambda}({_{\Lambda}^{6}{\rm H}})=5$ MeV, 
the average of 4.2 and 5.8 MeV predicted in Refs.~\cite{DLS63,Aka99}, 
respectively, with respect to $^{5}{\rm H} + \Lambda$. This choice is not 
critical, since $T_{\rm sum}$ varies merely by 50 keV upon varying 
$B_{\Lambda}({_{\Lambda}^{6}{\rm H}})$ by 1 MeV, negligibly low with respect 
to the experimental energy resolutions $\sigma_{T}(\pi^{+})=0.96$ MeV and 
$\sigma_{T}(\pi^{-})=0.84$ MeV for $p_{\pi^{+}}\approx 250$ MeV/c and 
$p_{\pi^{-}}\approx 130$ MeV/c. Therefore, the FINUDA energy resolution for 
a $\pi^{\pm}$ pair in coincidence is $\sigma_{T}=1.28$ MeV. Evaluating the 
r.h.s. of Eq.~(\ref{Ebal}) one obtains $T_{\rm sum}=203\pm 1.3$ MeV for 
$^{6}_{\Lambda}$H candidate events. In practice we have focused on events in 
the interval $T_{\rm sum}=203\pm 1$ MeV, corresponding to only $\sim 77\%$ of 
the FINUDA total energy resolution; this value was chosen as a compromise 
between seeking to reduce contamination from background reactions discussed 
in more detail below, and maintaining reasonable statistics, which resulted 
in a somewhat narrower interval than the experimental resolution. The raw 
spectrum of $T_{\rm sum}$ for $\pi^{\pm}$ pair coincidence events is shown 
in Fig.~\ref{fig3}, where events satisfying $T_{\rm sum}=203\pm 1$ MeV are 
indicated by a vertical (red) bar. 

\begin{figure*}[htbp] 
\vspace{-5mm} 
\begin{center} 
\includegraphics[width=150mm]{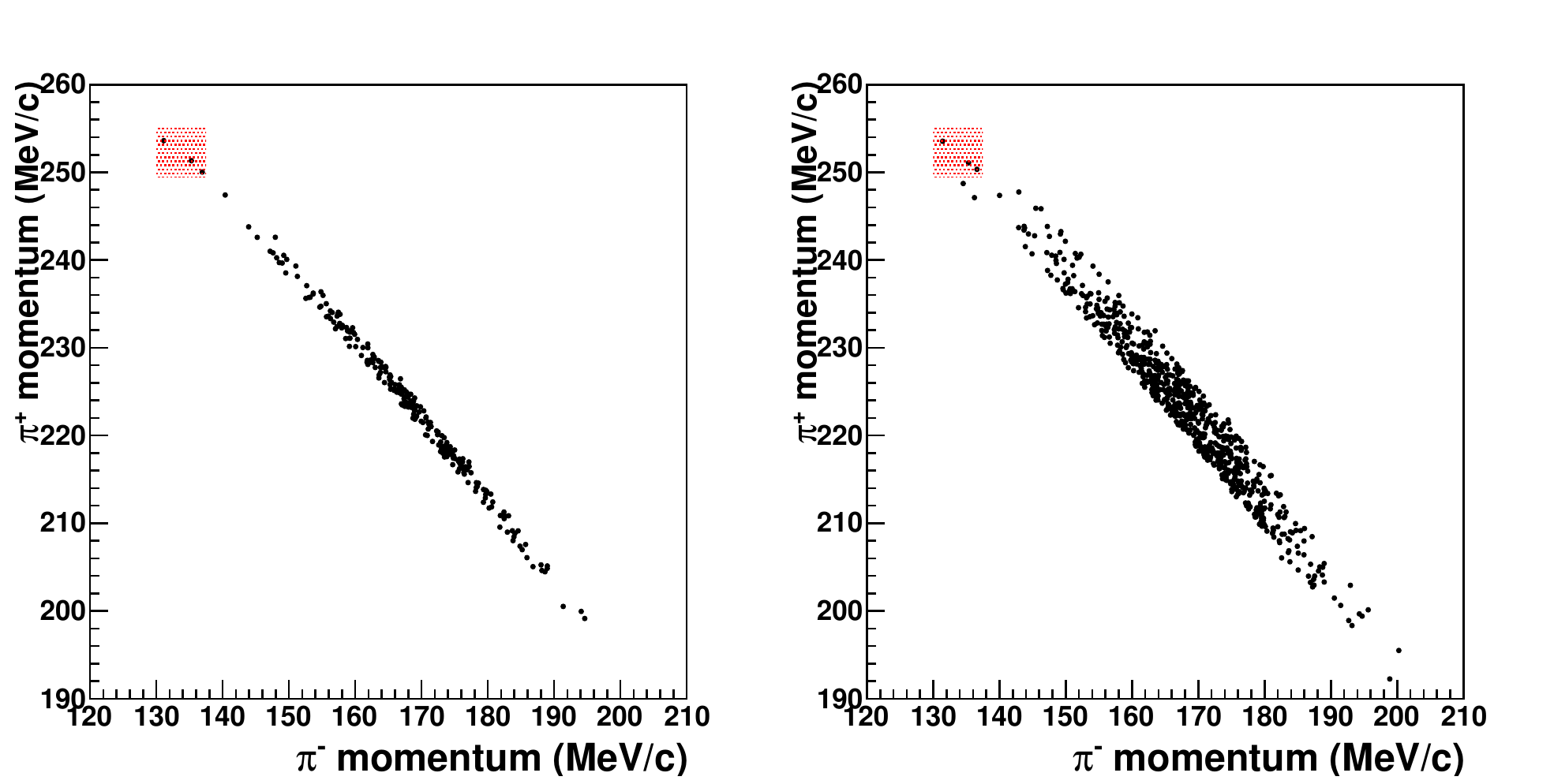} 
\caption{(color online). $\pi^{+}$ momentum vs $\pi^{-}$ momentum for 
$^{6}$Li target events with $T_{\rm sum}=202-204$ MeV (l.h.s.) and with 
$T_{\rm sum}=200-206$ MeV (r.h.s.). The shaded (red) rectangle on each 
side consists of a subset of events with $p_{\pi^{+}}=250-255$ MeV/c and 
$p_{\pi^{-}}=130-137$ MeV/c.} 
\label{fig4} 
\end{center} 
\vspace{-5mm} 
\end{figure*} 

Figure~\ref{fig4} (left) shows a 2-d plot in the $p_{\pi^{\pm}}$ plane 
of coincidence events selected in the band $T_{\rm sum}=202-204$ MeV. 
The distribution falls to zero at $p_{\pi^{+}}\simeq 245$ MeV/c and higher, 
and at $p_{\pi^{-}}\simeq 145$ MeV/c and lower. This is close to where 
$^{6}_{\Lambda}$H events are expected. Thus, to search for particle-stable 
$^{6}_{\Lambda}$H events below its $({^{4}_{\Lambda}{\rm H}}+2n)$ lowest 
threshold, using the two-body kinematics of Eqs.~(\ref{eq:production}) and 
(\ref{eq:decay}), a further requirement of $p_{\pi^{+}}>251.9$ MeV/c and 
$p_{\pi^{-}}<135.6$ MeV/c is necessary. In the final analysis we selected 
$p_{\pi^{+}}=(250-255)$ MeV/c and $p_{\pi^{-}}=(130-137)$ MeV/c, thus covering 
a $^{6}_{\Lambda}$H mass range from the $(\Lambda+{^{3}\mathrm{H}}+2n)$ 
threshold, about 2 MeV in the $^{6}_{\Lambda}$H continuum, down to 
a $^{6}_{\Lambda}$H bound somewhat stronger than predicted by Akaishi 
{\it et al.} \cite{Aka99}. This does not completely exclude eventual 
contributions from the production and decay of $({^{4}_{\Lambda}{\rm H}}+2n)$ 
as discussed below.

{\textbf{Results.}}~ 
Out of a total number of $\sim 2.7\cdot 10^{7}$ $K^-$ detected at stop in the 
$^{6}$Li targets, we found three events that satisfy the final requirements: 
$T_{\rm sum}=202-204$ MeV, $p_{\pi^{+}}=250-255$ MeV/c and $p_{\pi^{-}}=
130-137$ MeV/c, as shown within the shaded (red) rectangle in Fig.~2 left. 
Different choices of $T_{\rm sum}$ interval widths ($2-6$ MeV) and position 
(center in $202-204$ MeV), and of $p_{\pi^{\pm}}$ interval widths ($5-10$ and 
$8-15$ MeV/c respectively) with fixed limits at 250 and 137 MeV/c respectively 
to exclude the unbound region, do not affect the population of this selected 
rectangle. For example, no new candidate events appear in the shaded rectangle 
upon extending the cut $T_{\rm sum}=202-204$ MeV in the l.h.s. of Fig.~2 to 
$T_{\rm sum}=200-206$ MeV in the r.h.s. of the figure. 
A similar stability is not observed in the opposite corner of Fig.~2 where, 
on top of the events already there on the l.h.s., six additional events appear 
on the r.h.s. upon extending the $T_{\rm sum}$ cut of the l.h.s. 
Quantitatively, fitting the projected $\pi^{\pm}$ distributions of 
Fig.~2 left by gaussians, an excess of three events in both $p_{\pi^{\pm}}$ 
distributions is invariably found, corresponding to the shaded (red) 
rectangle. The probability for the three events to belong to the 
fitted gaussian distribution is less than $0.5\%$ in both cases. 
This rules out systematic errors associated with the 
present analysis selection. 

The three $^{6}_{\Lambda}$H candidate events are listed in Table~\ref{tab1} 
together with nuclear mass values derived separately from production 
(\ref{eq:production}) and from decay (\ref{eq:decay}). These mass values 
yield a mean value $M({^{6}_{\Lambda}{\rm H}})=5801.4\pm 1.1$ MeV, jointly 
from production and decay, where the error reflects the spread of the average 
mass values for the three events, and is larger than the 0.96 MeV and 0.84 MeV 
measurement uncertainties in production and decay, respectively, for each of 
the three events. We note that the mass value inferred from the third event by 
averaging on production and decay is about $2\sigma$ from the mean mass value, 
an observation that could indicate some irregularity in the reconstruction 
of the third event. To regain confidence, each one of the three events was 
checked visually for irregularities, but none was found. 

Furthermore, we note from Table~\ref{tab1} that the mass values 
associated with production are systematically higher than those associated 
with decay, by $0.98\pm 0.74$ MeV recalling the 1.28 MeV uncertainty for 
$T_{\rm sum}$ from which each of these mass differences is directly 
determined. Unlike the mean $^{6}_{\Lambda}{\rm H}$ mass value, the spread 
of the production vs decay mass differences is well within $1\sigma$. 
These mass differences are likely to be connected to the excitation spectrum 
of $_{\Lambda}^{6}{\rm H}$ as discussed below. 

\begin{table} 
\begin{center} 
\caption{Summed kinetic energy $T_{\rm sum}=T(\pi^{+})+T(\pi^{-})$, 
pion momenta $p_{\pi^{\pm}}$, and mass values inferred for the three 
$^{6}_{\Lambda}$H candidate events from production (\ref{eq:production}) 
and decay (\ref{eq:decay}). The mean mass value is $M({^{6}_{\Lambda}{\rm H}})
=5801.4\pm 1.1$ MeV, see text.} 
\label{tab1} 
\begin{tabular}{ccccc} 
\hline \hline 
\noalign{\smallskip} 
$T_{\rm sum}$  & $p_{\pi^{+}}$ & $p_{\pi^{-}}$ & 
$M(^{6}_{\Lambda}{\rm H})_{\rm prod.}$ & 
$M(^{6}_{\Lambda}{\rm H})_{\rm decay}$ \\ 
(MeV) & (MeV/c) & (MeV/c) &  (MeV) & (MeV) \\ 
\noalign{\smallskip} 
\hline 
\noalign{\smallskip} 
202.6$\pm$1.3 & 251.3$\pm$1.1 & 135.1$\pm$1.2 & 5802.33$\pm$0.96 & 
5801.41$\pm$0.84 \\ 
202.7$\pm$1.3 & 250.1$\pm$1.1 & 136.9$\pm$1.2 & 5803.45$\pm$0.96 & 
5802.73$\pm$0.84 \\ 
202.1$\pm$1.3 & 253.8$\pm$1.1 & 131.2$\pm$1.2 & 5799.97$\pm$0.96 & 
5798.66$\pm$0.84 \\ 
\hline \hline 
\end{tabular} 
\end{center} 
\end{table}

{\textbf{Background estimate and production rate}}~ 
A complete simulation was performed of $K^{-}_{\rm stop}$ absorption reactions 
on single nucleons, as well as on correlated few-nucleon clusters, that lead 
to the formation and decay of $\Lambda$ and $\Sigma$ hyperons. Full details 
will be given elsewhere, here it is sufficient to focus on two chains of 
reactions likely to produce $\pi^{\pm}$ coincidences overlapping with those 
selected to satisfy $^{6}_{\Lambda}$H production (\ref{eq:production}) and 
decay (\ref{eq:decay}). 

(i) $\Sigma^+$ production 
\begin{equation} 
K^{-}_{\rm stop}+{^{6}{\rm Li}}\rightarrow\Sigma^{+}+{^{4}{\rm He}}+n+\pi^{-}, 
\label{bgd_S_1} 
\end{equation} 
where $p_{\pi^{-}}\leq 190$ MeV/c, followed by $\Sigma^{+}$ decay in flight 
\begin{equation} 
\Sigma^{+}\rightarrow n+\pi^{+} \ \ \ \ [p_{\pi^{+}}\leq 282\ \mathrm{MeV/c}]. 
\label{bgd_S_2} 
\end{equation} 
The $\Sigma^+$ production was treated in the quasi--free approach, following 
the analysis of the FINUDA experiment observing $\Sigma^{\pm}\pi^{\mp}$ pairs 
\cite{fnd_S}. This simulation is shown in Fig.~\ref{fig3}, normalized to the 
experimental distribution area. It provides too sharp a decrease in the 
200-210 MeV region. To have a satisfactory description a contribution 
($\sim 25\%$) from a pure 4-body phase space mechanism was added and 
a fair agreement was obtained ($\chi^{2}=40/39$) in the 180-220 MeV range. 
The simulated background spectra reproduce reasonably the projected 
distributions of $\pi^{\pm}$ momentum too, showing in particular only 
a small contribution to the signal region, evaluated to be $0.16\pm 0.07$ 
expected events (BGD1). 

(ii) $^{4}_{\Lambda}$H production 
\begin{equation} 
K^{-}_{\rm stop}+{^{6}{\rm Li}}\rightarrow {^{4}_{\Lambda}{\rm H}}+2n+\pi^{+}, 
\label{bgd_4LH_1} 
\end{equation} 
where $p_{\pi^{+}}\leq 252$ MeV/c, with $^{4}_{\Lambda}$H decay at rest 
\begin{equation} 
^{4}_{\Lambda}{\rm H} \rightarrow {^{4}{\rm He}} + \pi^{-} \ \ \ \ 
[p_{\pi^{-}}\sim 132.8\ \mathrm{MeV/c}]. 
\label{bgd_4LH_2} 
\end{equation} 
The $\pi^{-}$ momentum in this $^{4}_{\Lambda}$H decay is close to 
$p_{\pi^{-}}\sim 134$ MeV/c from the two-body decay of $^{6}_{\Lambda}$H, 
evaluated assuming $B_{\Lambda}({^{6}_{\Lambda}{\rm H}})=5$ MeV as 
discussed above. A value of $0.04\pm 0.01$ expected events for the 
(\ref{bgd_4LH_1})-(\ref{bgd_4LH_2}) reaction chain, negligible when 
compared to BGD1, was obtained under most pessimistic assumptions for 
the various terms of the calculation. 

All other reaction chains that could produce $\pi^{\pm}$ coincidences within 
the described selection ranges were ruled out by the selections applied. 
Turning to potential instrumental backgrounds, we note that these could result 
from fake tracks, misidentified as true events by the track reconstruction 
algorithms. To this end we considered, with the same cuts, events coming from 
different nuclear targets used in the same runs ($^{7}$Li, $^{9}$Be, $^{13}$C, 
$^{16}$O). We found one event coming from $^{9}$Be. Furthermore, we considered 
events relative to the $^{6}$Li targets, selected with a value of $T_{\rm sum} 
=193-199$ MeV, so as to search for neutron-rich hypernuclei produced on 
the other targets. No event was found. We evaluate $0.27\pm 0.27$ expected 
fake events from $^{6}$Li, due to instrumental background (BGD2). 

To recap, the estimated number of events due to physical and instrumental 
backgrounds feeding through the selection criteria are $0.16\pm 0.07$ (BGD1) 
and $0.27\pm 0.27$ (BGD2), giving a total background of $0.43\pm 0.28$ 
expected events. Thus, using Poisson distribution, the three 
$^{6}_{\Lambda}$H-assigned events do not arise from background to a confidence 
level of 99$\%$. 
The statistical significance of the result is S=7.1 considering only the physical 
background, S=3.9 considering both physical and instrumental backgrounds.

Given the above background estimates, plus efficiency, target purity and cut 
estimates, it is possible to evaluate the product $R(\pi^+)\cdot {\rm BR} 
(\pi^{-})$, where $R(\pi^+)$ is the $^{6}_{\Lambda}$H production rate per 
$K^{-}_{\rm stop}$ in reaction (\ref{eq:production}) and ${\rm BR}(\pi^-)$ 
the branching ratio for the two-body $\pi^-$ decay (\ref{eq:decay}): 
\begin{equation} 
R(\pi^+)\cdot {\rm BR}(\pi^{-})=(2.9\pm 2.0)\cdot 10^{-6}/K^{-}_{\rm stop}. 
\label{rate} 
\end{equation} 
Details will be given in a separate report. 
Assuming BR$(\pi^{-})=49\%$, as for the analogous 
$^{4}_{\Lambda}{\rm H}\rightarrow {^{4}{\rm He}}+\pi^{-}$ decay \cite{KEK89}, 
we find $R(\pi^+)=(5.9\pm 4.0)\cdot 10^{-6}/K^{-}_{\rm stop}$, 
fully consistent with the previous FINUDA upper limit \cite{FIN06}. 

\begin{figure}[htbp] 
\begin{center} 
\includegraphics[width=80mm]{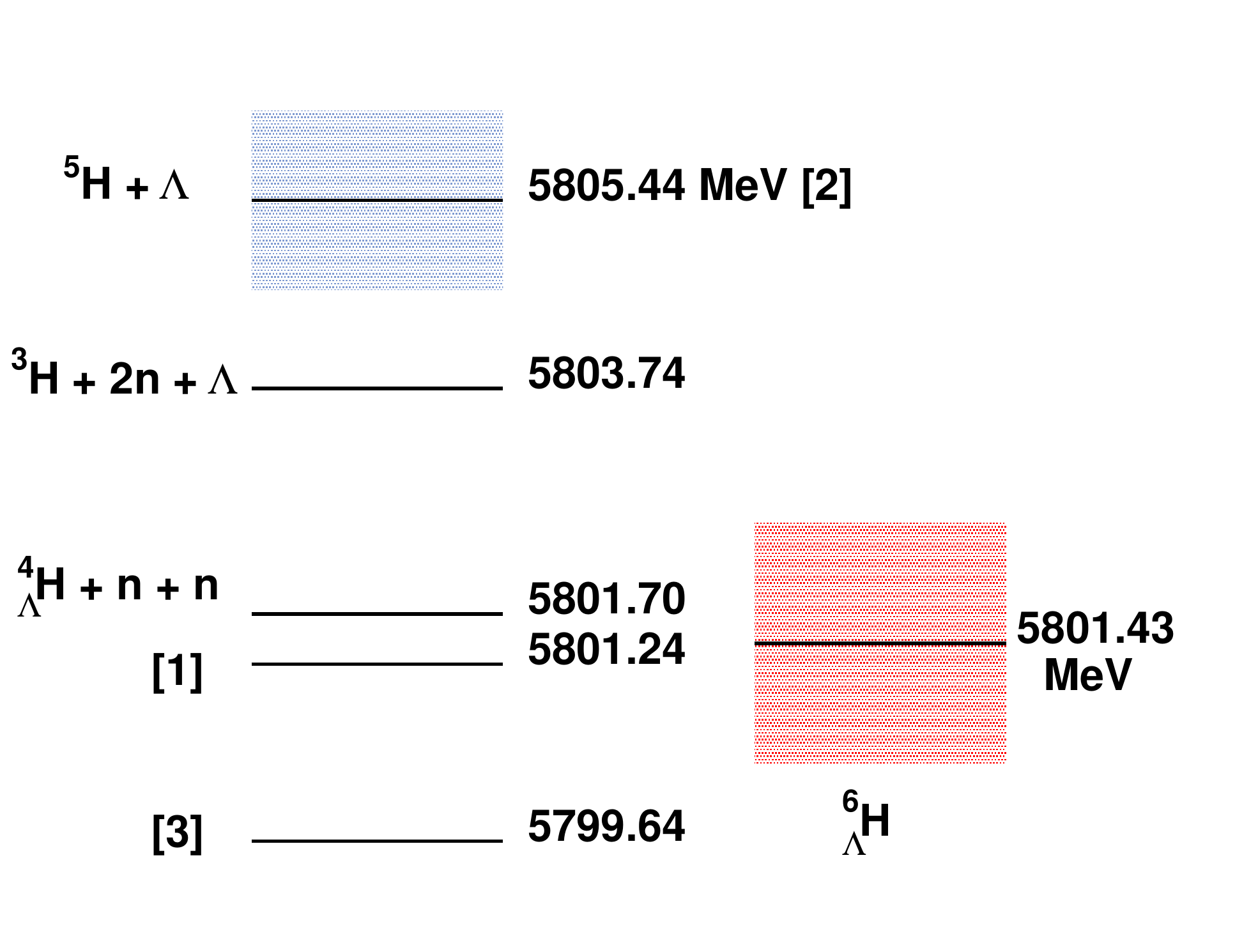}  
\caption{(color online). $^{6}_{\Lambda}$H mass (r.h.s.) from three 
$^{6}_{\Lambda}$H candidate events, as related to several particle 
stability thresholds and theoretical predictions (l.h.s.).} 
\label{fig5} 
\end{center} 
\end{figure}

{\textbf{Discussion and Conclusion.}}~ 
Table~\ref{tab1} yields a mean value 
$B_{\Lambda}({_{\Lambda}^{6}{\rm H}})=4.0\pm 1.1$ MeV with respect 
to $^{5}{\rm H}+\Lambda$, as shown in Fig.~\ref{fig5}, in good 
agreement with the estimate 4.2 MeV \cite{DLS63} and close to 
$B_{\Lambda}({_{\Lambda}^{6}{\rm He}})=4.18\pm 0.10$ MeV 
(with respect to $^{5}{\rm He}+\Lambda$) for the other known $A=6$ 
hypernucleus \cite{Davis05}, but considerably short of Akaishi's prediction 
$B_{\Lambda}^{\rm th}({_{\Lambda}^{6}{\rm H}})=5.8$ MeV \cite{Aka99}. 
This indicates that coherent $\Lambda N-\Sigma N$ mixing in the $s$-shell 
hypernucleus $_{\Lambda}^{4}{\rm H}$ \cite{Aka00} becomes rather ineffective 
for the excess $p$ shell neutrons in $_{\Lambda}^{6}{\rm H}$. Indeed, recent 
shell-model calculations by Millener indicate that $\Lambda N-\Sigma N$ mixing 
contributions to $B_{\Lambda}$ and to doublet spin splittings in the 
$p$ shell are rather small, about $(10\pm 5)\%$ of their contribution 
in $_{\Lambda}^{4}{\rm H}$ \cite{Mil10}. 

Next, we ask whether the three events that give evidence for a particle-stable 
$_{\Lambda}^{6}{\rm H}$ provide additional information on its excitation 
spectrum which is expected to consist of a $0^+$ g.s. and $1^+$ excited state 
as in $_{\Lambda}^{4}{\rm H}$ (1.04 MeV), and a $2^+$ excited state as for 
the $p$-shell dineutron system in $^{6}{\rm He}$ (1.80 MeV). In fact, it 
is $_{\Lambda}^{6}{\rm H}(1^+)$ that is likely to be produced in reaction 
(\ref{eq:production}) simply because Pauli spin is conserved in production 
at rest, and the Pauli spin of $^{6}$Li is $S=1$ to better than $98\%$ 
\cite{Mil10}. The weak decay (\ref{eq:decay}), however, occurs from 
$_{\Lambda}^{6}{\rm H}(0^+)$ g.s. since the (unseen) $\gamma$ transition 
$1^+ \to 0^+$ is about three orders of magnitude faster than weak decay. 
Indeed, the production vs decay mass difference $0.98\pm 0.74$ MeV 
extracted from the three $_{\Lambda}^{6}{\rm H}$ events listed in 
Table~\ref{tab1} is comparable to the underlying 1.04 MeV $1^+$ 
excitation in $_{\Lambda}^{4}{\rm H}$ but, again, smaller than 
the 2.4 MeV predicted by Akaishi {\it et al.} \cite{Aka99}. 
If this is the case, then the $B_{\Lambda}$ value for the g.s. would 
be larger by 0.5 MeV than that determined above, amounting to 
$B_{\Lambda}({^{6}_{\Lambda}{\rm H_{\rm g.s.}}})=4.5\pm 1.2$ MeV. 
This scenario requires further exploration, experimental as well as 
theoretical. 

In conclusion, we have presented the first evidence for heavy hyper-hydrogen 
$_{\Lambda}^{6}{\rm H}$, based on detecting three events shown to be clear 
of instrumental and/or physical backgrounds. The derived binding energy 
of $_{\Lambda}^{6}{\rm H}$ limits the strength of the coherent 
$\Lambda N-\Sigma N$ mixing effect predicted in neutron-rich strange matter 
\cite{Aka99} and together with the conjectured $0^+ - 1^+$ doublet splitting 
it places a limit on this mixing that could orient further explorations of 
other neutron-rich hypernuclei. 
A search of $_{\Lambda}^{6}{\rm H}$ and $_{~\Lambda}^{10}{\rm Li}$ in the 
$(\pi^-,K^+)$ reaction at 1.2 GeV/c on $^6$Li and $^{10}$B, respectively, 
is scheduled in the near future at J-PARC.


\begin{thebibliography}{99} 

\bibitem{DLS63} R.H. Dalitz and R. Levi-Setti, Nuovo Cimento \textbf{30}, 489 
(1963); L. Majling, Nucl. Phys. A \textbf{585}, 211c (1995). 

\bibitem{Kor01} A.A. Korsheninnikov {\it et al.}, Phys. Rev. Lett. 
\textbf{87}, 092501 (2001). 

\bibitem{Aka99} Y.~Akaishi and T.~Yamazaki, in {\it Frascati Physics Series} 
\textbf{XVI}, 59 (1999); S.~Shinmura, K.S.~Myint, T.~Harada, and Y.~Akaishi, 
J. Phys. G \textbf{28}, L1 (2002); 
For a recent review, see Y.~Akaishi and K.S.~Myint, AIP Conf. Proc. 
\textbf{1011}, 277 (2008), Y.~Akaishi, Prog. Theor. Phys. Suppl. \textbf{186}, 
378 (2010), and references therein. 

\bibitem{Sch08} J. Schaffner-Bielich, Nucl. Phys. A \textbf{804}, 309 (2008), 
Nucl. Phys. A \textbf{835}, 279 (2010). 

\bibitem{FIN06} M. Agnello {\it et al.}, Phys. Lett. B \textbf{640}, 145 
(2006), including a detailed description of the FINUDA experiment, 
in particular for the $(K^{-}_{\rm stop},\pi^+)$ production reaction. 




\bibitem{KEK89} H.~Tamura {\it et al.}, Phys. Rev. C \textbf{40}, R479 (1989). 

\bibitem{spectrFND} M.~Agnello {\it et al.}, Phys. Lett. B \textbf{698}, 
219 (2011). 

\bibitem{fnd_S} M.~Agnello {\it et al.}, Phys. Lett. B \textbf{704}, 474 
(2011). 

\bibitem{Davis05} D.H. Davis, Nucl. Phys. A \textbf{754}, 3 (2005). 

\bibitem{Aka00} Y. Akaishi, T. Harada, S. Shinmura, and K.S. Myint, 
Phys. Rev. Lett. \textbf{84}, 3539 (2000). 

\bibitem{Mil10} D.J. Millener, Lect. Notes Phys. \textbf{724}, 31 (2007), 
and references therein. 

\end{thebibliography}
\end{document}